\documentclass[a4paper]{article}

\usepackage{INTERSPEECH2019}

\usepackage{multirow}

\title{An Investigation Into On-device Personalization of End-to-end Automatic Speech Recognition Models}
\name{Khe Chai Sim, Petr Zadrazil, Fran{\c{c}}oise Beaufays}
\address{Google, USA}
\email{\{khechai,binus,fsb\}@google.com}

\begin{document}

\maketitle
\begin{abstract}
Speaker-independent speech recognition systems trained with data from many users are generally robust against speaker variability and work well for a large population of speakers.
However, these systems do not always generalize well for users with very different speech characteristics. This issue can be addressed by building personalized systems that are designed to work well for each specific user.
In this paper, we investigate the idea of securely training personalized end-to-end speech recognition models on mobile devices so that user data and models never leave the device and are never stored on a server. We study how the mobile training environment impacts performance by simulating on-device data consumption.
We conduct experiments using data collected from speech impaired users for personalization.
Our results show that personalization achieved 63.7\% relative word error rate reduction when trained in a server environment and 58.1\% in a mobile environment. 
Moving to on-device personalization resulted in 18.7\% performance degradation, in exchange for improved scalability and data privacy. To train the model on device, we split the gradient computation into two and achieved 45\% memory reduction at the expense of 42\% increase in training time. 
\end{abstract}

\section{Introduction}

Automatic speech recognition (ASR) systems are trained with thousands of hours of speech data from many users 
to be robust against speaker variability and achieve state-of-the-art performance~\cite{li2017acoustic,soltau2017neural,narayanan2018toward}.
However, these systems do not work well for users whose voice characteristics are different from the main training population.
A common solution to this problem is to adapt a well-trained model using a small amount of user-specific data to build a personalized model for each user~\cite{sim2017_springer,li2010comparison, saon2013speaker,senior2014improving,samarakoon2016factorized,tan2015cluster,wu2015multi,swietojanski2014learning,xue2014singular,zhao2016low}.

In order to personalize server-side ASR models, we would need to build infrastructure to train, evaluate and maintain a model per user on the server.
Recent work~\cite{he2019streaming} shows that an offline ASR system based on RNN-T can run on mobile devices and achieve state-of-the-art performance for Google's voice search traffic.
This opens a new possibility for on-device personalization of ASR models, where personalized models are trained on users' devices.
Not only does this let us avoid having to store and host personalized models on a server, it also allows us to complete all modelling tasks (data acquisition, training, evaluation, and inference) directly on device without having to send personal data to a server.
This is related to federated learning~\cite{mcmahan2017communication,konevcny2016federated}, where model updates are computed locally across multiple devices and securely aggregated on a server to train a shared model.
Federated learning has been successfully applied to mobile keyboard prediction and query suggestions~\cite{hard2018federated,yang2018applied}.
There are many challenges pertaining to on-device training of ASR models due to limited memory and storage.
We did experiments on a dataset of speech impaired users
to investigate how training in a mobile environment impacts ASR performance compared to a regular server-side training.
Specifically, we simulated on-device data availability and consumption while fine-tuning different sub-parts of the full RNN-T to understand the best way to personalize ASR RNN-T models.

The remainder of this paper is organized as follows.
Section~\ref{sec:model} presents the on-device RNN-T speech recognition model architecture.
Section~\ref{sec:personalization} describes on-device personalization and discusses several strategies to cope with limited storage and memory in
a mobile training environment. 
%We also describe a sliding window approach to simulate training data consumption on mobile devices.
Section~\ref{sec:experiments} presents experimental results showing how various training conditions affect the performance of the personalized ASR models.

\section{On-device ASR Model Architecture}
\label{sec:model}

\begin{figure}[t]
\centering
\includegraphics[width=0.38\textwidth]{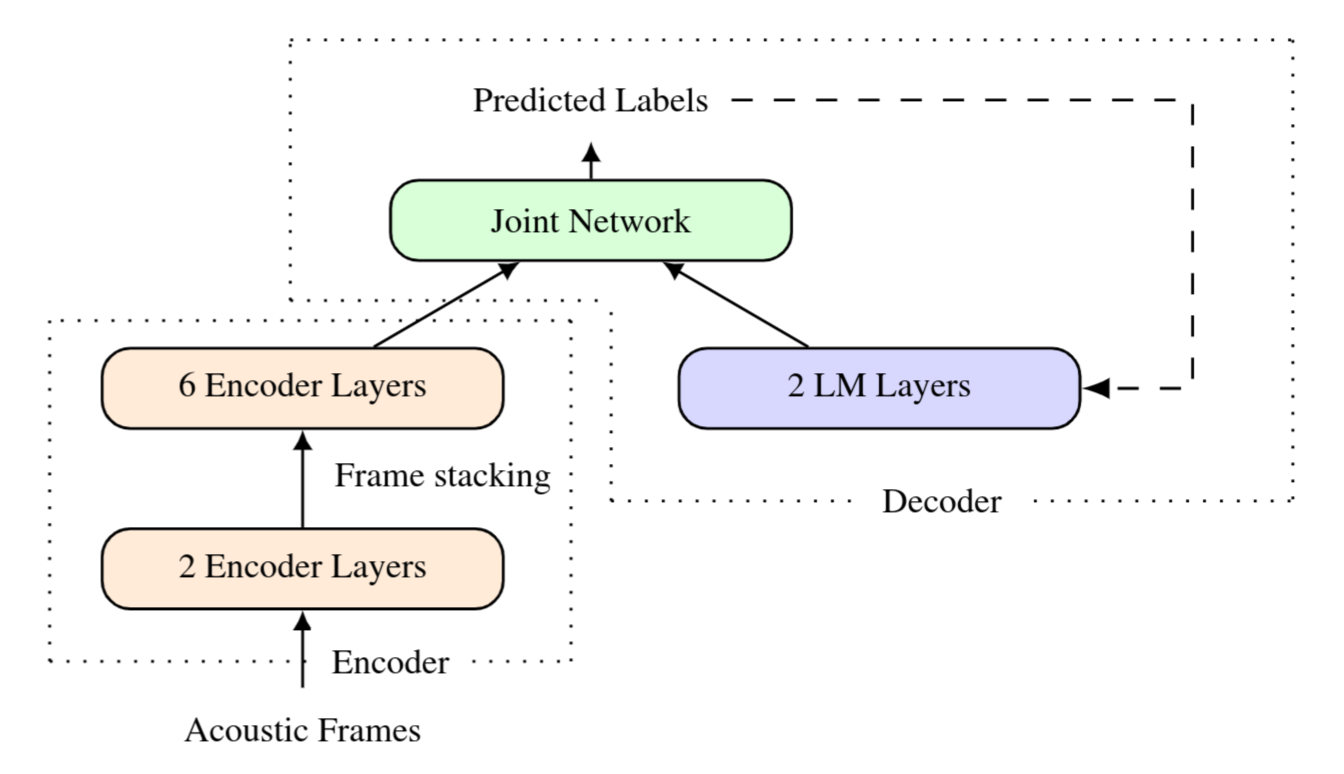}
\caption{Architecture of the on-device RNN-T model (8 encoder layers, 2 LM layers and a joint network with 1 hidden layer).}
\label{fig:rnnt_arch}
\end{figure}

Recently, the recurrent neural network transducer (RNN-T)~\cite{graves2012sequence} and Listen-Attend-Spell (LAS)~\cite{chan2016listen} end-to-end ASR models
have been successfully applied to large vocabulary continuous speech recognition~\cite{rao2017rnnt,chiu2018state}.
Moreover, a fast and accurate offline ASR system based on the RNN-T has been successfully deployed on mobile devices~\cite{he2019streaming}.
Fig.~\ref{fig:rnnt_arch} depicts the on-device RNN-T model architecture. It consists of
an encoder with 8 Long Short Term Memory (LSTM)~\cite{sak2014long} layers for the acoustic features, an encoder with 2 LSTM layers for the label sequences (denoted as the language model (LM) component) and a joint network with a single hidden layer. The LM and joint network components are referred to as the decoder.
Each LSTM layer has 2048 hidden units along with a projection to 640 units.
3 consecutive frames of 80-dimensional log Mel features (extracted every 10 milliseconds) are stacked together to form the 240-dimensional inputs to the network~\cite{pundak2016lfr}.
There is also frame stacking after the second encoder layer with a stride size of 2 to increase the frame rate to 60 milliseconds.
The network outputs correspond to 75 graphemes and a blank symbol~\cite{graves2012sequence}.

\section{On-device Personalization}
\label{sec:personalization}

The goal of ASR model personalization is to adapt a speaker-independent model to work well for specific users. 
Due to the proliferation of mobile devices, it is important to be able to train personalized ASR models on a mobile device.
Previously, personalization of ASR models on mobile device was achieved by adapting or biasing the language model using contextual information ({\em e.g} using contact information from on-device address books)~\cite{mcgraw2016personalized}.
Recently, an RNN-T ASR model has been developed for mobile devices and achieves state-of-the-art performance on a test set of Google's voice search traffic~\cite{he2019streaming}.
In that work, personalization was achived by fusing a contextual language model with RNN-T.
In this paper, we focus on building personalized RNN-T model which are trained on mobile devices.
In the following, we will discuss several aspects on how to cope with limited memory and storage in a mobile training environment.

\subsection{Data Storage}
\label{sec:sliding_window}

For on-device personalization, speech data is acquired and stored on device in a {\em training cache}. 
Unlike server-side training, it is not possible to store a large amount of data on a mobile device. It is also a good practice to store data only for a limited period of time to ensure that private user data is always being cycled and that only relevant recent data is available for training. Therefore, it is not possible to reuse the training data indefinitely.
\begin{figure}[t]
\centering
\includegraphics[width=0.3\textwidth]{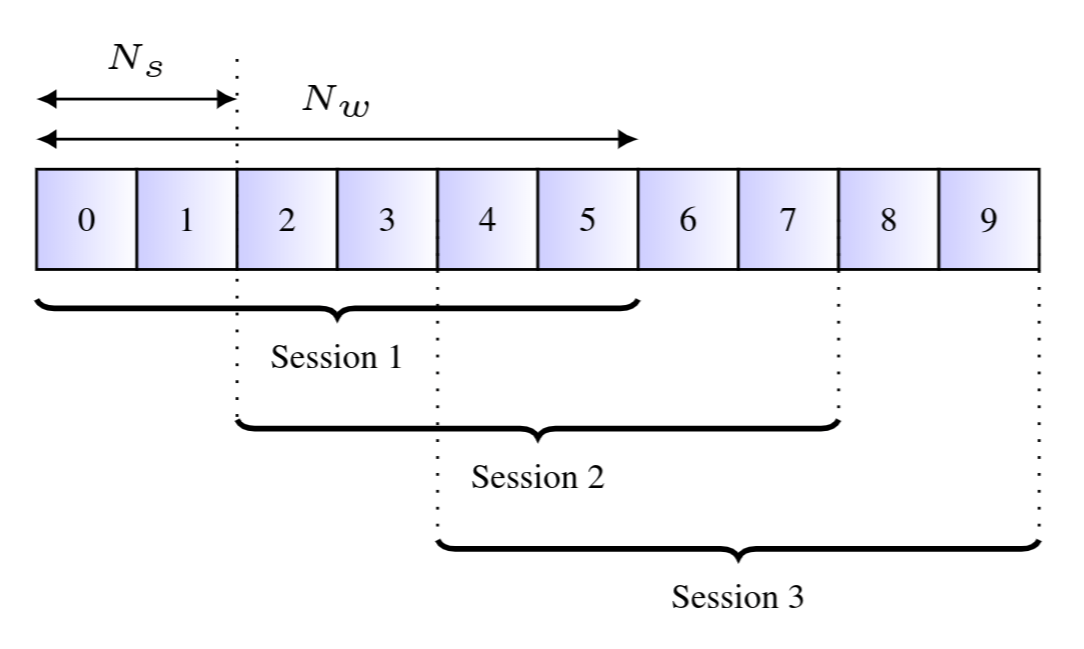}
\caption{Illustration of data windowing to simulate on-device data consumption.}
\label{fig:windowing}
\end{figure}
A {\em sliding window} concept can be used to simulate the constraints on how data is consumed on a mobile device. Training is performed over multiple sessions, where each session sees a window of training data. The window size ($N_w$) simulates the size of the training cache (the maximum number of examples to be stored on device at a given time). The window shift ($N_s$) determines the number of {\em new examples} in each session, which reflects the data acquisition rate (usage).
An example is illustrated in Fig.~\ref{fig:windowing}, with $N_w{=}6$ and $N_s = 2$. 
\begin{table}[t]
  \caption{Examples seen per mini-batch ($N_w{=}6$, $N_s{=}2$, $B{=}3$, $E_s{=}2$).}
  \label{tab:windowed_data}
  \centering
  \begin{tabular}{cccc}
    \toprule
    \textbf{Session} & \textbf{Epoch} & \textbf{Mini-batch} & \textbf{Data} \\
    \midrule
    1 & 1 & 1 & {\tt \{0, 1, 2\}} \\
    1 & 1 & 2 & {\tt \{3, 4, 5\}} \\
    1 & 2 & 1 & {\tt \{0, 1, 2\}} \\
    1 & 2 & 2 & {\tt \{3, 4, 5\}} \\
    \hline
    2 & 1 & 1 & {\tt \{2, 3, 4\}} \\
    2 & 1 & 2 & {\tt \{5, 6, 7\}} \\
    2 & 2 & 1 & {\tt \{2, 3, 4\}} \\
    2 & 2 & 2 & {\tt \{5, 6, 7\}} \\
    \hline
    3 & 1 & 1 & {\tt \{4, 5, 6\}} \\
    3 & 1 & 2 & {\tt \{7, 8, 9\}} \\
    3 & 2 & 1 & {\tt \{4, 5, 6\}} \\
    3 & 2 & 2 & {\tt \{7, 8, 9\}} \\
    \bottomrule
  \end{tabular}
\end{table}
For each session, training is performed in a regular way where training data is partitioned into mini batches of size $B$ and with multiple epochs per session ($E_s$). 
Note that $E_s$ is different from the {\em effective training epoch} (given by ${(E_s \times N_w)}/{N_s}$), which is 
the total number of times a training sample is used for updating the model.
This is because the same example is reused across multiple training sessions.
Table~\ref{tab:windowed_data} shows how data is consumed within each training session with
$B{=}3$ and $E_s{=}2$. 

\subsection{Memory}

The baseline RNN-T model (as described in Section~\ref{sec:model}) has approximately 117M parameters.
Once the model is trained on a server, the weight matrices are quantized to 8-bit integers for deployment. 
The resulting inference model is faster and more memory efficient.
However, training this model requires significantly more memory to facilitate gradient computation.
It is not possible to train the entire model with full precision on current high end mobile devices.
Rather than reducing memory consumption during training by building a smaller end-to-end model, we considered the following two approaches to reduce memory consumption. This allowed us to preserve the baseline state-of-the-art performance of the RNN-T.

\subsubsection{Selected Layers}
\label{sec:partial}

Training memory can be reduced by freezing the earlier layers of the encoder network\footnote{Note that for gradient computation, derivatives are back-propagated from the output all the way back to the earliest layer to be updated. Therefore, it is better to freeze the earlier layers.}.
Consequently, outputs from the frozen layers can be treated as static pre-computed features, which can be computed efficiently using the compressed inference model.
\begin{table}[t]
  \caption{Number of parameters for different parts of the model.}
  \label{tab:params}
  \centering
  \begin{tabular}{lrr}
    \toprule
    \textbf{Model} & \textbf{Number of parameters} & \textbf{Percentage} \\
    \midrule
        Joint & 901k & 0.8 \\
        LM & 19M & 16.6 \\
        Decoder & 20M & 17.4 \\
        Encoder 7 & 12M & 10.1 \\
        Encoder 6--7 & 24M & 20.2 \\
        Encoder 5--7 & 35M & 30.3 \\
        Encoder 4--7 & 47M & 40.5 \\
        Encoder 3--7 & 59M & 50.6 \\
        Encoder 2--7 & 76M & 65.2 \\
        Encoder 1--7 & 88M & 75.3 \\
        Encoder 0--7 & 96M & 82.6 \\
        All & 117M & 100.0 \\
    \bottomrule
  \end{tabular}
\end{table}
Table~\ref{tab:params} shows the breakdown of the number of parameters for different parts of the model. The majority of the model parameters come from the acoustic encoder stack (82.6\%). The decoder component only contributes 17.4\% of the model parameters, most of which are from the 2 LM encoder layers. 
Table~\ref{tab:params} also reports the cumulative sum of the parameters of the encoder layers, starting from the last layer. For example, `Encoder 4--7' refers to the total number of parameters for encoder layers 4 through 7. We were able to train on device, a subset of the model including the last 4 encoder layers (Encoder 4--7) and the decoder without other memory reduction techniques.

\subsubsection{Memory Efficient Gradient Computation}
\label{sec:split_gradient}

\begin{figure}
\centering
\includegraphics[width=0.4\textwidth]{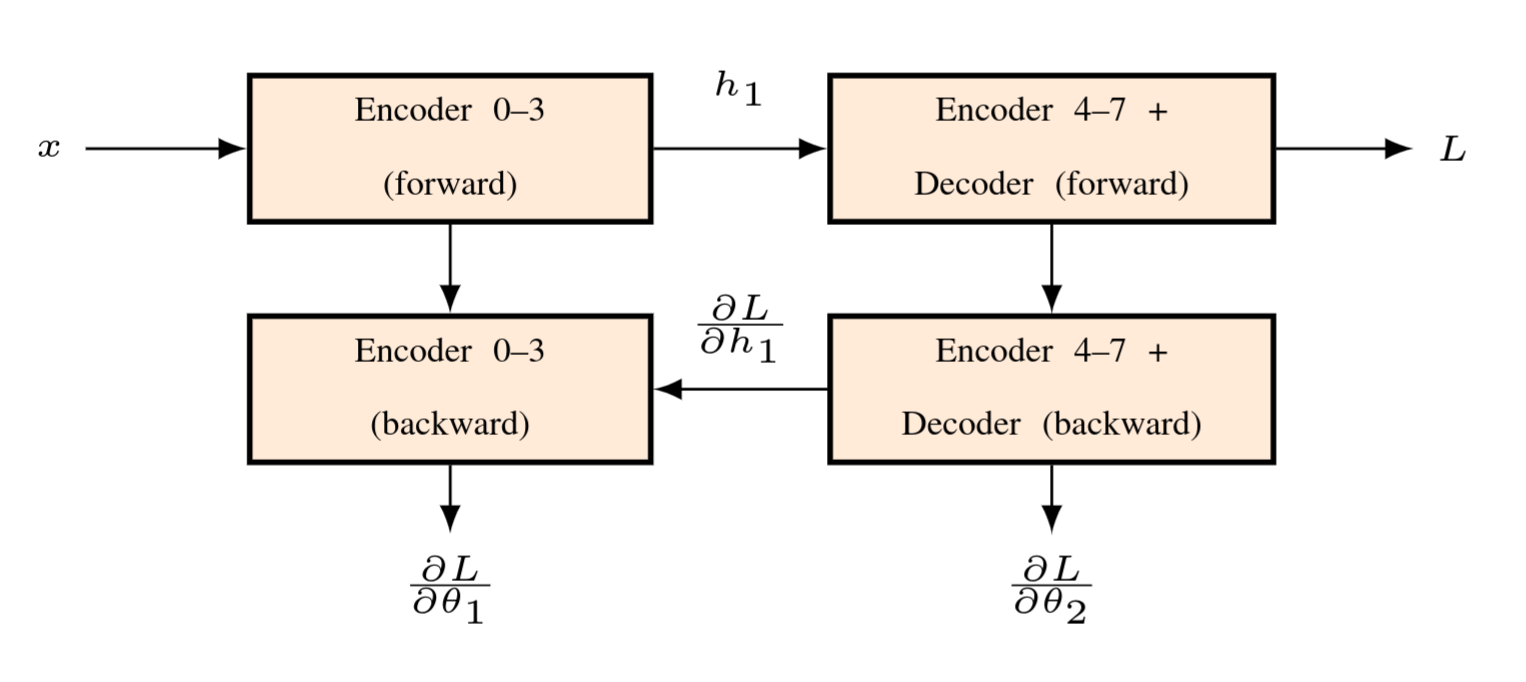}
\caption{Block diagram of split gradient computation.}
\label{fig:split_grad}
\end{figure}

Alternatively, the training graph can be split into smaller sub-graphs such that gradients are computed separately for each sub-graph. This significantly reduces the memory footprint at the expense of the additional loading and saving of the parameters for each sub-graph.
Fig.~\ref{fig:split_grad} illustrates the sequential computation of the split gradients. The model is partitioned into 2 sub-graphs. The first sub-graph consists of the first 4 encoder layers (Encoder 0--3) while the second sub-graph consists of the last 4 encoder layers and the decoder (Encoder 4--7 + Decoder). 
For each training step, the forward pass for the first sub-graph is computed. 
Then, the forward and backward passes of the second sub-graph are computed to obtain ${\partial L}/{\partial \theta_2}$ and ${\partial L}/{\partial h_1}$. 
Finally, ${\partial L}/{\partial \theta_1}$ is obtained by recomputing the forward pass and then computing the backward pass for the first sub-graph. 
$\theta_1$ and $\theta_2$ refer to the parameters associated with the first and second sub-graphs respectively.

\section{Experimental Results}
\label{sec:experiments}

In this section, we present experimental results from personalizing the speech RNN-T using the sliding window approach described in Section~\ref{sec:sliding_window} to simulate data consumption on mobile devices. All models are trained and evaluated using Tensorflow~\cite{abadi2016tensorflow}. The RNN-T loss and gradients are computed using the efficient implementation as described in~\cite{bagby2018efficient}.

We use a dataset consisting of English speech data collected from 20 amyotrophic lateral sclerosis (ALS) patients. These patients have speech impairments with varying degrees of severity.
\begin{table}[t]
    \centering
    \caption{Dataset.}
    \label{tab:dataset}
    \begin{tabular}{l|rr}
    \toprule
        \textbf{Model} & \textbf{Train} & \textbf{Test} \\
    \midrule
        Number of utterances & 24,188 & 2,000 \\
        Duration of speech (hours) & 30.6 & 2.5 \\
    \bottomrule
    \end{tabular}
\end{table}
Table~\ref{tab:dataset} summarizes the amount of train and test data used for experiments.
Each speaker has an average of 1.5 hours of training data. See~\cite{hassidim2019} for more details about the data.
\begin{table}[t]
    \centering
    \caption{Performance of baseline RNN-T model.}
    \label{tab:baseline}
    \begin{tabular}{lr}
    \toprule
        \textbf{Dataset} & \textbf{WER} \\
    \midrule
        Voice Search & 7.3 \\
        ALS & 35.6 \\
    \bottomrule
    \end{tabular}
\end{table}
We used the RNN-T model described in Section~\ref{sec:model} as our baseline.
This model was trained with 35 million anonymized hand-transcribed English utterances (∼27,500 hours), from Google's voice search traffic~\cite{he2019streaming}.
Table~\ref{tab:baseline} shows the performance of the baseline model on 2 test sets.
The model achieved state-of-the-art word error rate (WER) performance of 7.3\% on Google's voice search test set. On the other hand, the WER performance 
on the ALS dataset is 35.6\%, which is significantly worse, due likely to the drastically different acoustic characteristics between speakers in the ALS dataset and the training population.

\begin{table}[th]
    \centering
    \caption{Word error rate comparison for different models.}
    \label{tab:wer_vs_models}
    \begin{tabular}{l|rr}
    \toprule
        \textbf{Model} & \textbf{WER} & \textbf{Rel. Imp.} \\
    \midrule
        Baseline & 35.6 & --- \\\hline
        Joint & 30.8 & 13.6 \\
        LM & 32.5 & 8.8 \\
        Decoder & 31.8 & 10.9 \\
        Encoder 7 & 29.4 & 17.6 \\
        Encoder 6--7 & 27.1 & 23.8 \\
        Encoder 5--7 & 25.6 & 28.1 \\
        Encoder 4--7 & 22.0 & 38.4 \\
        Encoder 3--7 & 20.8 & 41.7 \\
        Encoder 2--7 & 20.0 & 44.0 \\
        Encoder 1--7 & 18.9 & 46.9 \\
        Encoder 0--7 & 19.6 & 44.9 \\
        All & 22.3 & 37.4 \\
    \bottomrule
    \end{tabular}
\end{table}
First, we evaluated the personalization performance in a mobile environment, using the following configuration as our baseline setting to simulate the way data is consumed on mobile devices: $N_w{=}100$, $N_s{=}4$, $B{=}10$ and $E_s{=}2$. We compared the performance of fine-tuning all or parts of the model. In all cases, we used momentum optimizer with a learning rate of $10^{-4}$. The word error rate performance is summarized in Table~\ref{tab:wer_vs_models}. 
We observe that fine-tuning only the last 7 encoder layers (1 -- 7) yields the best performance of 18.9\%, or 46.9\% relative improvement over the baseline model. 
Moreover, it is possible to fine-tune the entire model or a large number of parameters without running into over-fitting issue. This is inline with the results reported for domain adaptation~\cite{sim2018domain}.
Fine-tuning the LM layers gave the worse performance. In fact, fine-tuning only the joint network achieved better performance compared to the LM layers (30.8\% vs. 32.5\%), despite having 20 times fewer parameters ({\em c.f.} Table~\ref{tab:params}).
Similarly, it is better to fine-tune only the encoder rather than the entire model. 
This is not surprising because the LM layers learn only from the label sequences, making it difficult to adapt to variations in acoustic features between speakers. 

\begin{table}[t]
    \centering
    \caption{Comparison of word error rate performance with and without on-device constraints.}
    \label{tab:unconstrained_vs_constrained}
    \begin{tabular}{l|ccc}
    \toprule
        \textbf{Model} & \textbf{Unconstrained} & \textbf{Constrained} & \textbf{Change (\%)} \\
    \midrule
        Joint & 24.1 & 30.8 & -27.7 \\
        Decoder & 25.3 & 32.5 & -28.6 \\
        Encoder & 12.8 & 19.6 & -53.7 \\
        All & 14.9 & 22.3 & -49.7 \\
    \bottomrule
    \end{tabular}
\end{table}
In Table~\ref{tab:unconstrained_vs_constrained}, we compare the performance between training in a server environment (unconstrained) versus a mobile environment (constrained). The former does not use the windowing strategy while the latter uses the same windowing configuration as above. In both cases, the effective training epoch is 50~\footnote{Increasing training epochs gave marginal further improvements.}. 
From the simulation, we observe that moving from server training to on-device training resulted in 27.7\%--53.7\% relative increase in WER. This is a rather substantial degradation. In the subsequent experiments, we investigate how different mobile training conditions will impact personalization performance.

\begin{table}[th]
    \centering
    \caption{Comparison of word error rate performance of adapting all 8 encoder layers by varying $N_w$ ($N_s{=}4$ and $E_s{=}2$).}
    \label{tab:wer_vs_cache_size}
    \begin{tabular}{c|c|rr}
    \toprule
        \textbf{$N_w$} & \textbf{Effective Epoch} & \textbf{WER} & \textbf{Rel. Imp} \\
    \midrule
        100 & 50 & 19.6 & --- \\
        200 & 100 & 18.2 & 7.5 \\
        300 & 150 & 17.2 & 12.6 \\
        400 & 200 & 16.0 & 18.7 \\
        500 & 250 & 15.3 & 22.0 \\
    \bottomrule
    \end{tabular}
\end{table}
Table~\ref{tab:wer_vs_cache_size} compares the WER perfomance of adapting all 8 encoder layers by varying $N_w$. Increasing $N_w$ will increase the training cache size, and therefore increases the number of sessions in which each training utterance will be used. As a result, the effective epoch is also increased. We observe that increasing $N_w$ from 100 to 500 consistently improved the WER performance from 19.6\% to 15.3\% (22.0\% relative improvement).
To determine if this improvement is due to an increase in the effective epoch, we increased $E_s$ from 2 to 4, 6 and 8 -- and found no improvement
\begin{table}[th]
    \centering
    \caption{Comparison of word error rate performance of adapting all 8 encoder layers by varying both $N_w$ and $N_s$ ($E_s{=}2$).}
    \label{tab:wer_vs_num_utts}
    \begin{tabular}{c|c|rr}
    \toprule
        \textbf{$N_w$} & \textbf{$N_s$} & \textbf{WER} & \textbf{Rel. Imp} \\
    \midrule
        200 & 4 & 18.2 & --- \\
        300 & 6 & 17.2 & 5.3 \\
        400 & 8 & 15.8 & 13.3 \\
        500 & 10 & 15.2 & 16.8 \\
    \bottomrule
    \end{tabular}
\end{table}
We also conducted another set of experiments where the effective epoch is kept constant at 100 by varying both $N_w$ and $N_s$. The results, as shown in Table~\ref{tab:wer_vs_num_utts}, are similar to those in Table~\ref{tab:wer_vs_cache_size}. 
Each utterance is used the same number of times in all the 4 settings in Table~\ref{tab:wer_vs_num_utts}, but in a different order. 
A larger window corresponds to a longer interval at which an utterance is reused. This is apparently crucial to achieve better improvement. 
Note that when the window size is large enough to cover the entire training set, we are effectively training in the unconstrained environment that is used is regular server-side training of RNNT models.
To summarize, we found that updating encoder layers 1--7 using $N_w{=}500$ and $N_s{=}10$ achieved the best WER performance of 14.8\% (58.1\% relative improvement over the baseline). 

\begin{figure}[ht]
    \centering
    \includegraphics[width=0.35\textwidth]{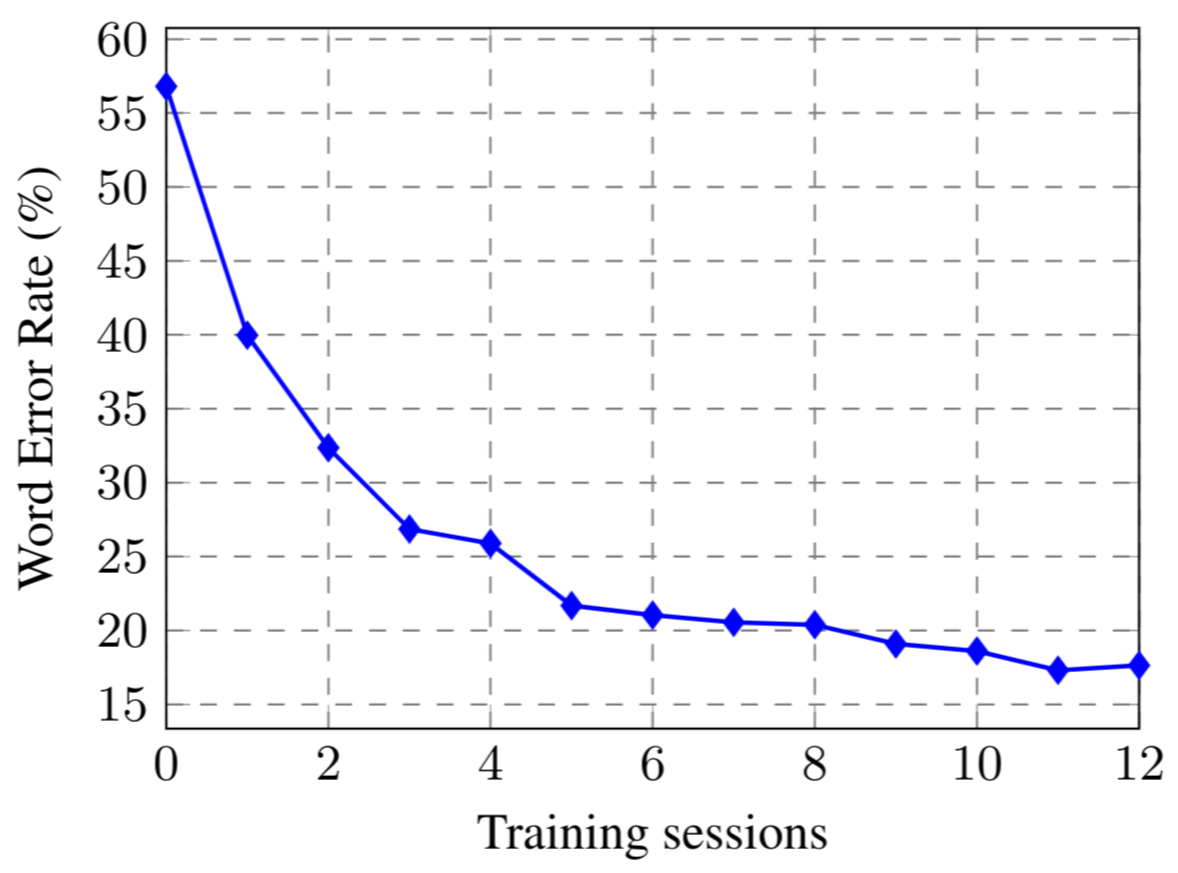}
    \caption{WER performance of updating encoder layers 1--7 versus training sessions ($N_w{=}500$, $N_s{=}10$ and $E_s{=}2$).}
    \label{fig:wer_vs_steps}
\end{figure}
With the sliding window setup, it is easy to study how the WER performance changes as training examples are progressively consumed.
Fig.~\ref{fig:wer_vs_steps} shows the WER trend versus training sessions for one user using the best personalized model. The baseline model achieved 56.8\% WER. After one training session (500 utterances), the WER performance dropped to 40.0\%. After 11 training sessions (600 utterances\footnote{Each subsequent training session consumes 10 new utterances.}),
the model achieved the best performance of 17.3\%, which is a 70\% relative improvement over the baseline model.

\begin{table}[t]
    \centering
    \caption{Memory and training speed on mobile device.}
    \label{tab:speed}
    \begin{tabular}{c|c|rr}
    \toprule
        \multirow{2}*{\textbf{Model}}
        & \textbf{Gradient} & \textbf{Memory} & \textbf{Time Per} \\
        & \textbf{Computation} & \textbf{(MB)} & \textbf{Utterance} \\
    \midrule
        Encoder (1024) & Combined & 1,187 & 2.6s \\
        Encoder (1024) & Split & 649 & 3.7s \\
        Encoder (2048) & Split & 1,497 & 12.5s \\
    \bottomrule
    \end{tabular}
\end{table}
Finally, we ran benchmark experiments on a Pixel 3 device to get realistic measures of on-device training speed.
Table~\ref{tab:speed} reports the memory usage and the training time to process one utterance.
We choose a smaller model as our baseline by reducing the number of hidden units of all the LSTM layers from 2048 to 1024 and the projection dimension from 640 to 320. This provides us a comparison point for split gradient computations runs with the full size model (as described in Section~\ref{sec:split_gradient}).
For the smaller network, split gradient computation reduces the memory usage from 1,187MB to 649MB (45\% reduction), at the expense of 42\% increase in the training time (2.6s to 3.7s per utterance).
The larger network used 1,497MB memory and takes 12.5s to process one utterance. This translates to about 3.5 hours per training session ($N_w=500$ and $B=10$).

\section{Conclusions}
\label{sec:conclusions}

In this paper, we proposed using on-device personalization to improve the performance of
automatic speech recognition. We conducted experiments to investigate how various on-device training conditions impact the speech recognition performance. Specifically, we use a sliding window model to simulate the way data is consumed in a mobile training environment. 
To train the model on device, one viable approach to reducing memory consumption is splitting the gradient computation in parts.
We found that, given an RNN-T model that is well-trained for ordinary users, it is possible to fine-tune the encoder layers to achieve significant performance improvements for speech impaired users, using an average of 1.5 hours of training data per user. 
Although we observed 18.7\% relative increase in WER as a result of moving from server to local training,
on-device personalization offers a more scalable and secure solution that does not require personal data and models to be sent to a server.
Obtaining high quality transcriptions for on-device learning remains a challenge which we will address in our future work.
\section{Acknowledgement}

The authors would like to thank Yanzhang He for providing the baseline model for our experiments and Dhruv Guliani for proofreading the paper.

\bibliographystyle{IEEEtran}

\bibliography{references}

\end{document}